\documentclass[conference]{IEEEtran}
\IEEEoverridecommandlockouts
\usepackage{cite}
\usepackage{amsmath,amssymb,amsfonts}
\usepackage{algorithmic}
\usepackage{graphicx}
\usepackage{textcomp}
\usepackage{xcolor}
\usepackage{hyperref, boolexpr, amsmath, float}
\def\BibTeX{{\rm B\kern-.05em{\sc i\kern-.025em b}\kern-.08em
    T\kern-.1667em\lower.7ex\hbox{E}\kern-.125emX}}

\begin{document}

\title{Blockchain Tree for eHealth}

\author{\IEEEauthorblockN{1\textsuperscript{st} Sergii Kushch}
\IEEEauthorblockA{\textit{Security and Trust research unit} \\
\textit{Bruno Kessler Foundation}\\
Trento, Italy \\
skushch@fbk.eu, kushch@yaros.co}
\and
\IEEEauthorblockN{2\textsuperscript{nd} Silvio Ranise}
\IEEEauthorblockA{\textit{Security and Trust research unit} \\
\textit{Bruno Kessler Foundation}\\
Trento, Italy \\
ranise@fbk.eu}
\and
\IEEEauthorblockN{3\textsuperscript{rd} Giada Sciarretta}
\IEEEauthorblockA{\textit{Security and Trust research unit} \\
\textit{Bruno Kessler Foundation}\\
Trento, Italy \\
giada.sciarretta@fbk.eu}
}

\maketitle

\begin{abstract}
The design of access control mechanisms for healthcare systems is challenging: it must strike the right balance between permissions and restrictions. In this work, we propose a novel approach that is based on the Blockchain technology for  storage patient medical data and create an audit logging system able to protect health data from unauthorized modification and access.  The proposed method consists of a tree structure: a main chain linked with the patient's identity and one or several subchains which are used for storing additional critical data (e.g., medical diagnoses or access logs).
\end{abstract}


\section{Introduction}
The design of access control mechanisms for healthcare systems is challenging. 
On the one hand, as these mechanisms  deal with sensitive data they must guarantee: 
\textsl{confidentiality}, in the sense that only the patient and doctors with specific 
access control policies and purposes can access the patient's personal health records (PHRs); 
and \textsl{integrity}, the PHR should not be modified without a clear evidence.
On the other hand, they should protect the safety of the patient, thus allowing doctors 
to access patients information quickly and without interruptions (e.g., in case of emergency). 
However, along with this flexible and frictionless access control comes the temptation 
of taking advantage of it. Indeed, as reported in~\cite{Verizon24}, $58\%$ of incidents involved 
insiders, this makes healthcare the only industry in which insiders are the biggest 
threat to an organization.
The motives range from simple curiosity about a friend or family member, to the wish 
to damage a patient by revealing some sensitive data, or for money (e.g., receiving 
an insurance payments by using a stolen diagnosis). Thus, access control mechanisms 
in the healthcare context must strike the right balance between permissions and restrictions.

In Italy, many successful solutions have been developed from the regional healthcare 
systems to protect electronic PHRs combining access control mechanisms and 
authentication solutions with a high level of assurance on the user's identity 
(e.g., using smartcards). The problem is that still little has been done for auditing.  
Given the aforementioned issue with insiders, it is essential that all the access of an
healthcare organization should be tracked through a non repudiation logging system in a 
way to be able to attribute privilege abuses and deter employees from improper behaviors. 

To fill this gap, we propose to use of the Blockchain technology~\cite{Hyperledger12}, 
which is a well known technology used in Bitcoin~\cite{Debin06,Garay02,Matsuo03} and 
other cryptocurrencies~\cite{Bentov01,EOS11}. Successful attempts are also being made 
to introduce this technology in areas of bank transfers~\cite{Koles21}, logistic~\cite{DHL20}, 
energy~\cite{Kushch13}, IoT \cite{Fran14, Kushch22} and healthcare~\cite{Jiang23}.
In its essence, Blockchain is a distributed database in which each subsequent block 
is associated with the previous ones. 
The generation of each block must be confirmed by other participants using a so called 
\textsl{consensus algorithm} (e.g., Proof of Work (POW)~\cite{Jakobsson04,Laurie05}, 
Proof of Stake (POS) \cite{Buterin09,Chohan07}, Proof of Importance (POI) \cite{Beikverdi10}, 
Proof of Activity (POA) \cite{Bentov08}). 
It should be noted that, currently, the most used algorithm is POW. However, for our purposes, 
it is not suitable, since the generation of blocks, when using it, it is too expensive. 
In addition, for each transaction, it is necessary to pay to the miners who mine new blocks. 
For the healthcare systems, it would be optimal to use an algorithm in which the generation of 
blocks is as cheap as possible, and the transactions are free.

\begin{figure}[h]
\centering{\includegraphics[width=85mm]{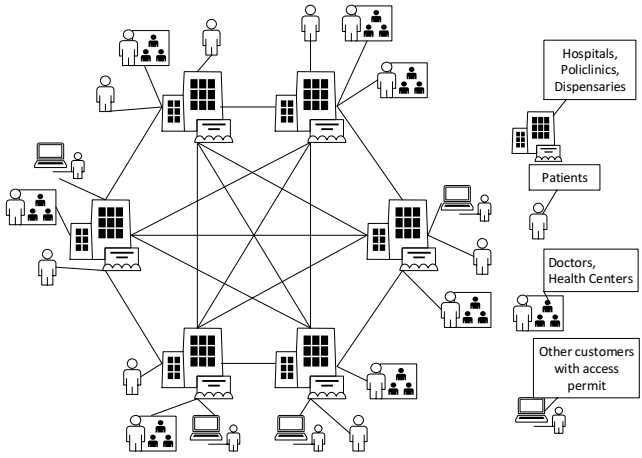}}
\caption{\textbf{The overall structure of the network. Node holders are hospitals, dispensaries, other medical organizations. The users are patients, doctors, as well as third parties authorized by the state or patient.}\label{figure1}}
\end{figure}

\paragraph{Use Case: Healthcare}
As shown in Figure~\ref{figure1}, the nodes of the considered network are servers of local branches of the healthcare system (hospitals, ambulatories or other medical organizations) that store personal information about citizens and electronic health records. The users of the information are patients, doctors, as well as third parties authorized by the state.

The network structure is a classical peer-to-peer (P2P) topology in which each element is connected to each one.
All nodes are equal. The network has a fixed number of nodes. Each node is verified and included in the list of approved nodes. This list is stored on each node and only devices from this list can create new blocks.

The main contribution of this paper is the presentation of a novel approach that is based on  the creation of a Blockchain tree, which is considered as a distributed storage of patient's PHR as well as an audit logging system able to protect PHRs from unauthorized modifications, and the use of the Proof of Authority (POA) consensus algorithm. In this case, all access attempts (successful and unsuccessful) will be stored in one of the side chains. This will not allow insiders 
to change logs about access (to hide privilege abuse) and any changes to PHRs without an authorized access.\\
\subsection{Related projects}
Currently there are several start-ups and commercial projects aimed at implementing Blokchein in health care. Developed on the Ethereum Blockchain, MedRec \cite{28MeRec} is a "system that gives priority to the patient agency by providing a transparent and accessible review of the history of the disease." MedRec is designed to store all patient information in one place, which makes it easier for patients and doctors to watching.
Connecting Care \cite{29ConCare} "uses care coordination and financial forecasting to help suppliers in integrated payments understand what happens to patients when they leave the hospital." It is currently on the market, helping health care providers determine how much they will be paying for patient care when included with multiple organizations.
The Thai Medical University Hospital and Digital Treasury \cite{30PHROS} recently released phrOS. It is aimed at increasing transparency between medical institutions by placing all the patient's medical information on the Blockchain.
Bloxine FarmaTrust \cite{31Farma} intends to help fight counterfeit drugs. The chain of supply visibility tracks drug changes or changes in any way. And, finally, the Consumer Confidence app allows customers to see their drug life cycle.
Electronic health records (EHR) \cite{32EHR} can be complex in management. The EHR of one healthcare provider for a patient may differ from another provider of the same patient. MTBC intends to change this with the help of application programming interfaces (APIs) and Blockchain. The idea is to pass control into the patient's hands. The patient will be able to decide whether to transfer records from one doctor to another. The blockchain API works on the Hyperledger platform and is currently available.
Hashed Health, a company focused on the development of Blockchain, focuses on healthcare, intends to make healthcare sector authorities more transparent and accessible. With Professional Credentials Exchange \cite{33HH}, chain members can check credentials and track records from various health professionals. This simplifies the process of hiring, and also provides the unchanging history of a professional medical career.
Change Healthcare develops a wide range of products focused on paying and managing data in the health sector. One of their latest developments \cite{34CHH} simplifies claims management and profitable cycle management. It helps hospitals and healthcare systems manage claims and money orders, improve patient fee collection, minimize bans and under-payments, and more effectively manage daily income and business cycles.
With MedicalChain \cite{35MCH} you get full access to and control over your personal medical data. Users can give doctors immediate access to their medical card through their mobile devices while they are stored in a reliable Blockhain. Patients can also wear bracelets that medical workers can scan to access a human's disease history if they are unconscious. He also offers telemedicine communication that allows online video consultation with doctors.\\
As can be seen from the list, existing projects are aimed at finance, medicines and information exchange between patients and medical institutions. In addition, most of the existing blocks are commercial projects for which every user has to pay money. Unlike a number of projects offered in this paper, the model involves the free participation of medical institutions, in addition, there will be no need to pay for each transaction. What is logical for such an important sphere as the provision of medical care. Another important point is the registration of access to user's personal information for a separate Blockchain. This will protect the system from hacking and, in the case of dissemination of confidential information, it will quickly find the culprit, as the system is protected from erasure or replacement of access logs.\\
{\color{blue}\paragraph{Paper Structure.}}
The paper is organized as follows. In Section~\ref{sec:background} we provide the background on Blockchain. Our approach, using Blockchain Tree, applied to the healthcare system is presented in Section~\ref{sec:approach}. 
Finally, we present the conclusions obtained from our research and discuss the 
possibilities for future work in Section~\ref{sec:conclusion}.

\section{Background}\label{sec:background}
Blocks in Blockchain are permanently recorded files that contain information about transactions of users. All transactions in the block are represented as strings in hexadecimal format (raw transaction format), which is hashed to obtain transaction identifiers (txid). On their basis, a hash of the block is built, which is taken into account by the subsequent block, ensuring the immutability and coherence of the registry. The unit hash value is compiled using the Merkle Tree, the concept of which was patented by Ralph Charles Merkle in 1979.

\begin{figure}[h]
\centering{\includegraphics[width=0.48\textwidth]{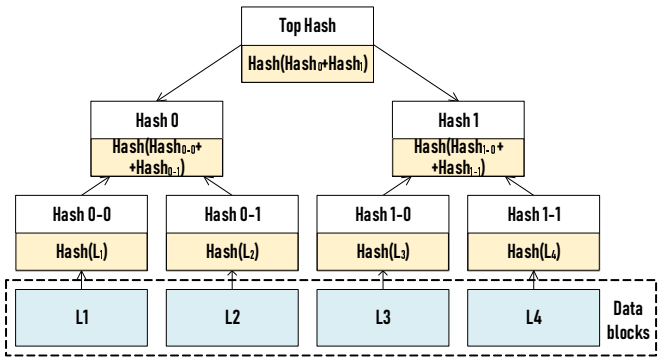}}
\caption{\textbf{An example of a binary hash tree~\cite{HashTree26}. Hashes 0-0 and 0-1 are the hash values of data blocks L1 and L2, respectively, and Hash 0 is the hash of the concatenation of Hashes 0-0 and 0-1.  }\label{figure2}}
\end{figure}

A Merkle Tree, or hash tree, is a binary tree whose leaf nodes are transaction hashes, and internal vertices are the results of the addition of the values of the associated vertices. The process is repeated until a single hash is obtained - the root of the Merkle tree (Merkle Root).
Figure~\ref{figure2} shows an example of a hash tree with four transaction-leaves L1, L2, L3, and L4. 
The construction of the tree is as follows: \emph{(i)} Hash 0-0, Hash 0-1, Hash 1-0, and Hash 1-1 are calculated as the hashes of the associated transactions (hash(L1), hash(L2), hash(L3), and hash(L4), respectively), \emph{(ii)} Hash 0, Hash 1 are calculated from the sum of transaction hashes (hash(Hash 0-0 + Hash 0-1), hash(Hash 1-0 + Hash 1-1), respectively), \emph{(iii)} finally the Top Hash is calculated as hash(Hash 0 + Hash 1). This can be generalized for a tree with $n$ leaves.
Since the Merkle tree is binary, the number of elements at each iteration must be even. Therefore, if a block contains an odd number of transactions, then the latter is duplicated and added to itself, e.g., hash(hash(L5) + hash(L5)).

In Blockchains, hash trees allow simplified verification of transactions: the verification of the integrity of the Blockchain entries is done by checking the hash blocks. Indeed, a main advantage is that clients willing to verify the integrity of data do not need to recalculate all the hashes to verify the transaction information, but they can ask for a Merkle’s evidence: it consists of the concatenation of the left and right hash of a branch, and validating the result against the parent. This step is repeated until the Merkle root is found. By adding the requested hashes and comparing them with the root, the client makes sure that the transaction is in its place.

This approach allows us  to work  with arbitrarily large amounts of data, since it significantly reduces the load on the network, since only the necessary hashes are downloaded. For example, the weight of a block with five maximum size transactions is more than 500 kilobytes. The weight of the proof of Merkle in the same case will not exceed 140 bytes.\\
A created block must be confirmed by more than 51\% of verified nodes. After that, the information is written to the block, added to the chain and sent to the other nodes. Thus, each element of the system stores a complete Blockchain. In the case of detecting a change in the information in an existing block, this block will be automatically replaced by the "right" that exists on at least 51\%, other nodes network.

Given that the blocks in the chain are sequentially linked to each other, changing of block will cause the whole chain to change. This will be detected and corrected by the rest of the network nodes.

\section{Main Results}\label{sec:approach}
The solution we offer consists of several interconnected parts. We propose to create a Blockchain, where each block contains information about one person (name, surname, date of birth, etc.) that allows you to uniquely identify him. In turn, each block is a “Genesis Block” for a Subchain, which contains a history of diseases for the same person. \\
Also, we propose to create a second Subchain to save user logs (Figure \ref{figure3}). In our opinion, it will help in the case of unauthorized access incidents investigating, since it will be impossible to change, or delete, the logs stored in the BC. Also, it will not be possible to change the stored information (diagnoses, prescribed treatment, etc.) for the purpose of insurance fraud or for other illegal purposes.

\subsection{Blockchain Structure}

\begin{figure}[t]
\centering{\includegraphics[width=85mm]{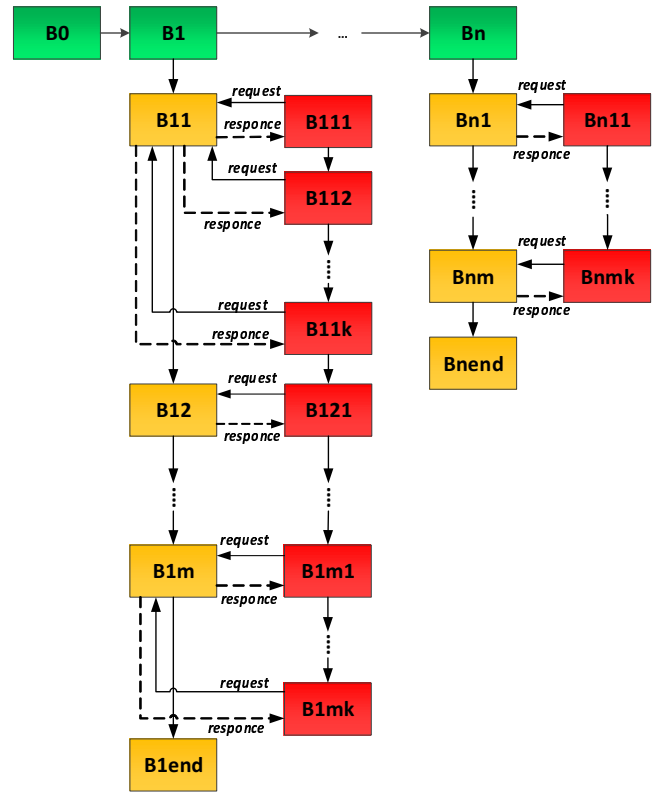}}
\caption{\textbf{Blockchain Tree: a system of three Blockchains that are connected with each other. The green chain is the main and contains personal information about patients; the yellow chain is the 1st Subchain and contains information about medical services, diseases etc.; the red chain is the 2nd Subchain and contains access logs of patients and medical staff.}\label{figure3}}
\end{figure}

As shown in Figure \ref{figure3}, the solution we propose consists of several interconnected parts organized as a Blockchain tree. Each main block (green color - B0, B1, ..., Bn) contains information about a patient (name, surname, date of birth, etc.) and is a "genesis block" of two Subchains. 

Subchain 1 (yellow) contains a history of the diseases for the related patient.  The blocks have two indices: 
	\begin{itemize}
			\item the 1st (from $1$ to $n$) is the block number in the main Blockchain and represents the patient identity;
			\item the 2nd (from $1$ to $m$)  is the block number in Subchain 1 and represents the medical record of the patient.
	\end{itemize}
	
For example $B_{23}$ is the third medical block of the second patient. It should be noted that the list of diagnoses, prescriptions, test results are constantly updated.  Doctors  should have access to both the latest results and the entire medical history. Therefore, each block in the chain stores a link to the previous block, recursively. Thus our solution permits a client to automatically assembles the last results, together with all the links to the history into a single document.\\
A level of protection is the ability to add information to the Blockchain only after validating the user's identity by entering the user's password or using another authentication method (e.g., using the electronic identity card), so we assume a private and permissioned blockchain.

Subchain 2 (red) contains user logs and consists of an additional index (from $1$ to $k$).
For example $B_{234}$ is the fourth log block of the third medical block of the second patient.
In our opinion, the adding of a block containing the logs will help investigation in the case of unauthorized access incidents, since the logs stored in a Blockchain will be impossible to change or delete.  Also, it will not be possible to change the stored information (diagnoses, prescribed treatment, etc.) for the purpose of insurance fraud or for other illegal purposes.

\begin{figure}[h]
\centering{\includegraphics[width=70mm]{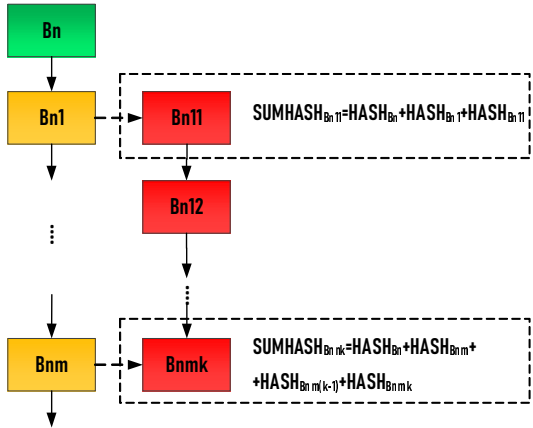}}
\caption{\textbf{Creating a cross HASH of the block of the second subchain.}\label{figure4_1}}
\end{figure}

After creating a new block in the yellow subchain, a new block is also created in the red subchain that contains three hashes one: of the main block which contains the basic information about a patient; of the new block of the yellow subchain; of the previous block of the red subchein. (Fig.\ref{figure4_1})
Thus, triple cross-reference to two Blockchains is performed, which significantly complicates the possibility of tamping with blocks or parts of a chain.

\begin{figure}[h]
\centering{\includegraphics[width=88mm]{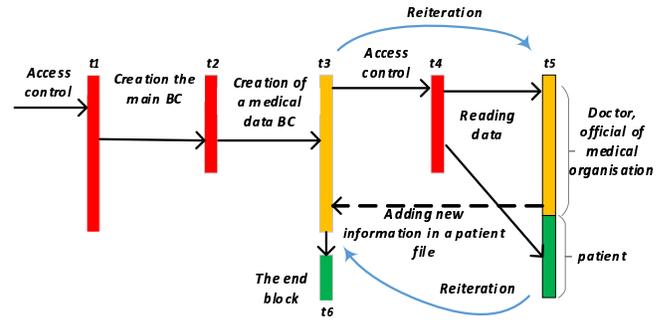}}
\caption{\textbf{Schematic of possible time steps of the proposed method use.}\label{figure5}}
\end{figure}

Thus, the system described above will have the following steps (Fig.\ref{figure5}) :
\textit{
   \begin{enumerate}
   \item The appeal of a citizen to the authorized department of the medical system and its registration in it after confirming the identity (before time t1);
    \item Creating a block in the main BC, which contains personal information of this person (time t1-t2);
    \item Creating a block 1 for the Subchain 1, for recording the future history of diseases (time t2-t3);
    \item Creating a block 1 for the Subchain 2, for recording access history to Subchain 1;
    \item When referring to a doctor, after validation of the card holder and the doctor, the blocks is automatically added to Subchain 1 and 2 (time t3);
    \item The patient's or an authorized medical officer's access to the patient's personal data - Subchain 1 (for example, to receive information about test results, appointments, etc.), after checking the access right, automatically creates a new block in Subchain 2. This block contains information about password holder (in case of unsuccessful entry - about access attempt), date, local time, place, what information was viewed, etc. (time t4-t6);
    \item If you need to add new patient information, the cycle repeats (the new block is created in Subchain 1);
    \item The Sabchain 1 Closing (time t6)
    \item After closing the subchain 1 there is possibility only for reading saved information (without adding new information). Subchain 2 continues to grow with each next access or access attempt.
   \end{enumerate}
}
Consider in more detail how this works.
There are different possible interactions with our system: \textsl{on-boarding of a new patient}, \textsl{writing new health data}, \textsl{reading an existing health data}. We assume that the user has undergone the procedure for obtaining valid access credentials (Fig.\ref{figure5}).

The on-boarding of a new patient causes the creation of the following blocks in our Blockchain tree:
\emph{(i)} a block containing personal information in the main Blockchain; \emph{(ii)} 
 a genesis block for Subchain 1; and
\emph{(iii)}  a genesis block for Subchain 2.

The writing of patient's health data by a medical officer causes the creation of a new block in Subchain 1 with the  automatic creation of the corresponding block in Subchain 2. The block in Subchain 1 contains the new health data, while the block in Subchain 2 contains log information about doctor's identity, date, local time, place, which information was added and so on. In case of unsuccessful entry, the new block contains the access attempt.\\
Instead, the reading of patient's health data in Subchain 1 (for example, to retrieve information about test results, appointments, etc.) automatically creates a new block only in Subchain 2. This block contains information about user's identity (patient or doctor), date, local time, place, which information was viewed and so on. In case of unsuccessful entry, the new block contains the access attempt.

\subsection{Further Considerations}
We report below some discussions on borderline cases and some extra features that should be considered.

\paragraph{The Change of Fiscal Code.}
The main identifier to which the chain is attached is a personal identifier. In Italy, this is the fiscal code. However, it should be noted that this code is tied to the name and surname, thus it will change if they changed. In this case, we will need to create an additional block in the Blockchain. This block will contain information about the new code, the old code, a link to the previous card and link to the rest of (the main) Blockchain, which existed before the name change. In this way, we avoid a Blockchain rupture. Repeated name changes can complicate the system, however, it will avoid fraud with the personality substitution. One of the ways to avoid this Blockchain complication is to introduce a single ID of a person identifier that does not change during life.

\paragraph{Stop Recording in Blockchain.}
 It is also necessary to envisage the situation when the patient leaves the citizenship or dies. In this case, it is necessary to block the possibility of further adding blocks to the patient's Subchain 1. After closing the Subchain 1 there is possibility only for reading saved information (without adding new information). Subchain 2 continues to grow with each next access or access attempt. This will provide additional protection for the system against possible fraud with bogus recipes and insurance. We propose to add a ``The Final Block" that marks the end of the Subchain and closes access to possibility to add any information. 

\begin{figure}[h]
\centering{\includegraphics[width=85mm]{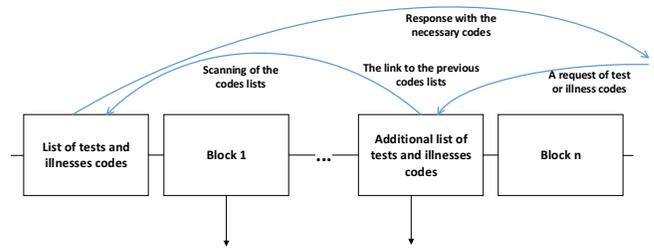}}
\caption{\textbf{Creation a list of medical tests and illnesses}\label{figure6}}
\end{figure}

\paragraph{The list of tests and illnesses.}
The main Blockchain contains a block with a list of all available tests and diseases (Fig. \ref{figure6}). In the case of the emergence of new diseases and tests, it is possible to update the list by creating an additional list in a new block. This block also contains a link to the main list.

\begin{figure}[h]
\centering{\includegraphics[width=85mm]{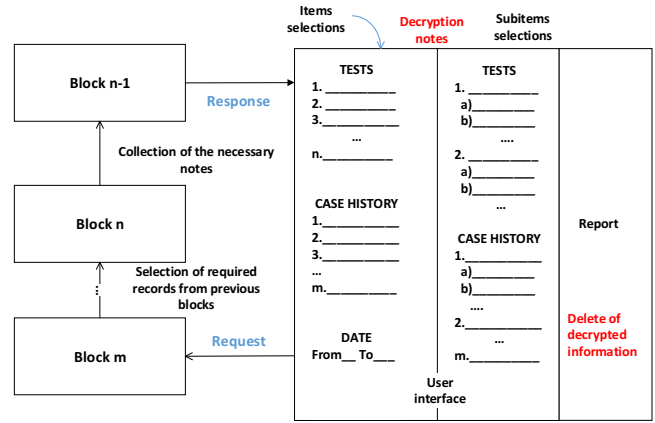}}
\caption{\textbf{Creation a report of medical tests and illnesses}\label{figure7}}
\end{figure}

\paragraph{Collection of an information from several blocks by user interface}
The block contains the latest test results and a link to previous test results of the same type. If the user needs to build a history of tests or a history of diseases - the records are read sequentially from all the blocks that contain them by clicking on the links.\\
The user interface (Fig. \ref{figure7}) reads the last block overhead. If the block contains the necessary record, it is added to the report, if not, the service record is read from the next block, etc.

\section{Summary and discussion}\label{sec:conclusion}
In this paper we propose a novel methodology based on Blockchain for building  storage, access control and document verification mechanisms in a healthcare. The proposed work is based on Subchains which are connected with the main Blockchain and with each other. The solution is more security than currently existing due to the mutual intersection of several Blockchain. This makes the process of hacking and spoofing critical information more difficult, since in the event of an attack, it will be necessary to change not one but several Blockchains, which considerably increases the cost of such an attack and makes it unprofitable for the attacker. The noted above methodology for building a storage system, access control and document verification can be used not only for medical-cards, but also for other documents, for example for ID-cards, driver's licenses, education documents, personal medical information and social security cards, etc. 
In addition, the proposed way to build a Blockchain allows you to create an arbitrary number of additional subchains and control access to information that they contain. For the health care system, this may be, for example, information on the availability of health insurance, which needs to be updated regularly.
Obviously, this work is only the first step in introducing the proposed concept into the system of preserving and protecting personal information in medicine, as well as to increase the level of access control there. The proposed methodology still needs further refinement in order to contribute to its reliable implementation and legal compliance (in particular referring to GDPR). For example, we neglected some of the problems of building real networks, such as delays in devices and communication lines. In future work, we will consider the problem of using various consensus algorithms with different types of Blockchain.

\bibliographystyle{IEEEtran}

\bibliography{main}

\end{document}